\documentclass[9pt,twocolumn,twoside]{opticajnl}
\journal{opticajournal} 

\setboolean{shortarticle}{true}

\usepackage{lineno}
\usepackage{amsmath}
\usepackage{float} 
\usepackage{placeins}

\title{Mitigating the phase-mismatch effect in non-resonant four-wave mixing enabled by optimal control
}

\author[1]{Ning Jia}
\author[2,*]{Hamid R. Hamedi}
\author[3,4,5,6 $\dagger$]{Jing Qian}

\affil[1]{Public Experiment Center, University of Shanghai for Science and Technology, Shanghai 200093, China}
\affil[2]{Institute of Theoretical Physics and Astronomy, Vilnius University, Saulėtekio 3, Vilnius LT-10257, Lithuania}
\affil[3]{School of Physics and Electronic Science, East China Normal University, Shanghai 200062, China}
\affil[4]{Chongqing Institute of East China Normal University, Chongqing 401120, People’s Republic of China}
\affil[5]{Shanghai Branch, Hefei National Laboratory, Shanghai 201315, China}
\affil[6]{Collaborative Innovation Center of Extreme Optics, Shanxi University, Taiyuan, Shanxi 030006, People’s Republic of China}

\affil[*]{hamid.hamedi@tfai.vu.lt}
\affil[$\dagger$]{jqian1982@gmail.com}

\begin{abstract}
Phase-mismatch in nonlinear optical processes can severely limit the propagation and conversion efficiency of light fields. Here, we present an efficient optimal-control strategy to mitigate the detrimental effects of phase-mismatch in an electromagnetically induced transparency medium via non-resonant four-wave mixing (FWM). By applying a set of fixed, linearly modulated coupling fields that induce a dark eigenmode, we globally optimize a single coupling detuning to minimize the spontaneous emission loss, the primary factor limiting conversion efficiency. Our approach outperforms existing FWM schemes by providing strong robustness against large phase-mismatch variations while maintaining efficient probe-to-signal conversion. These results offer a promising route toward more efficient nonlinear frequency conversion, alleviating the stringent requirement for phase matching in experiments.
\end{abstract}

\setboolean{displaycopyright}{false} 

\begin{document}

\maketitle


{\it Introduction.-} Phase mismatch, common in nonlinear atom-light systems, poses a major obstacle to efficient frequency conversion due to the limited coherence length of atomic media \cite{Harris:PRL2004,DELLANNO2006,meta-optics2021}. Except for elaborate experimental setups \cite{Scully:PRL1999,YFZhu:PRL2004,PRA023839,Wu:2022}, previous efforts to overcome this issue have shown that, for example, effective compensation using a small constant two-photon detuning can significantly suppress the spontaneous emission loss in a resonant FWM system \cite{Harris:PRL1996}. However, the required detuning must be precisely controlled, making the approach highly sensitive to technical imperfections. In parallel, an improved FWM scheme employing two spatially modulated coupling fields has been implemented to increase the conversion efficiency (CE) \cite{pra2018}. In this setup, the system remains in a transmission (dark) mode, preventing absorptive loss due to spontaneous emission. Nevertheless, the large phase-mismatch required by the ideal spatial-light-modulation condition ($\varphi\approx 90^{\circ}$) strongly disrupts the adiabaticity of the dark mode, posing a significant challenge to achieve a high CE under practical phase-mismatch conditions.

\begin{figure}[t]
    \centering
    \includegraphics[width=0.38\textwidth]{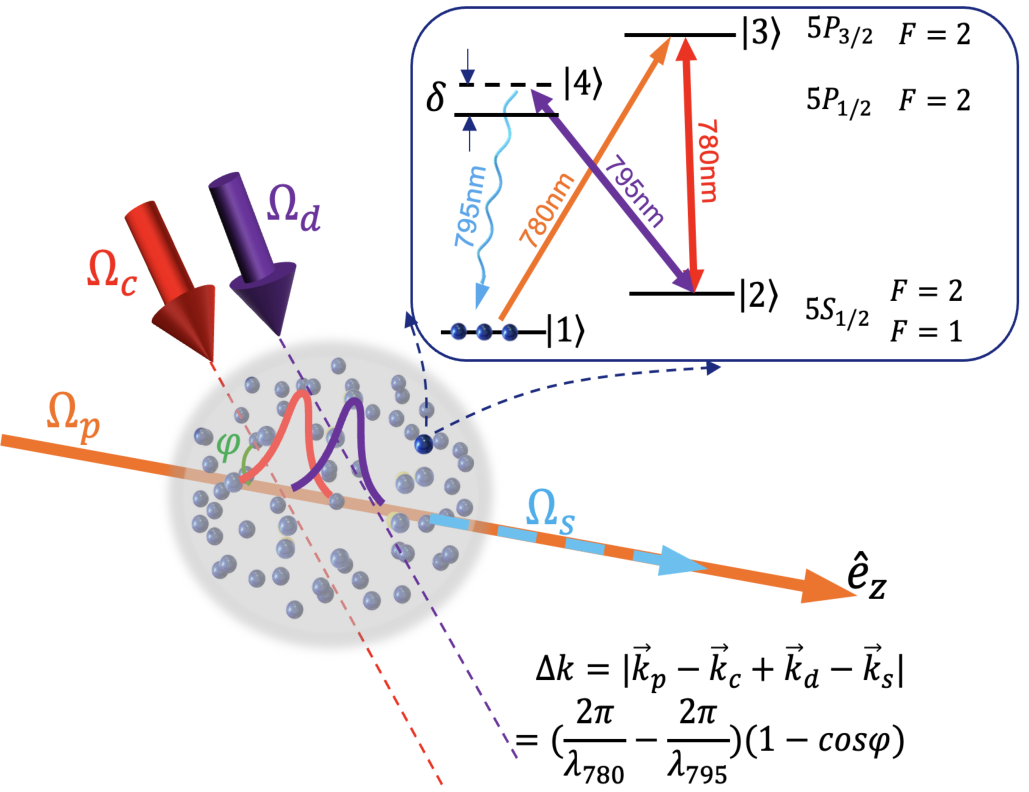} 
    \caption{\justifying Restriction by the phase-mismatch effect in a non-resonant double-$\Lambda$ FWM system \cite{PRA033845}. The phase-mismatch term $\Delta{k}$, arising from the separated angle $\varphi$ between the propagation directions of the probe ($\Omega_p$) and coupling ($\Omega_{d,c}$) beams is detrimental to the generation of signal field ($\Omega_s$). Box: Schematic level scheme and relevant energy states of cold $^{87}$Rb atoms utilized in this work.}
    \label{fig:model}
\end{figure}

In this work, we propose a global optimal-control method to mitigate the impact of phase-mismatch in a non-resonant FWM system, achieving a highly-efficient frequency conversion {{over a broad range of mismatch strengths}}. The numerical optimization is performed using Genetic Algorithm (GA) \cite{Jing:PRap}, with fine-tuning applied only to a single coupling detuning that varies smoothly in space, while maintaining a set of space-delayed coupling fields in an adiabatic regime to minimize spontaneous emission loss. The results explicitly demonstrate that a high-efficiency frequency conversion can be achieved
with enhanced robustness against significant mismatch errors. {{Notably, even modest phase mismatches—corresponding to small angles between the probe and coupling fields—can noticeably reduce CE in realistic experiments \cite{Wu:2022}. To illustrate the effectiveness and robustness of our approach, we explore mismatch angles up to $\varphi \approx 8^\circ$ for larger optical depth (OD), and extend to $\varphi \approx 13.6^\circ$ for moderate OD to intentionally stress-test our protocol. This work therefore extends the scope of phase-mismatch control by combining it with an advanced optimal-control strategy, which differs from the approach introduced in Ref.~\cite{ol2020}, and holds great promise for applications in versatile scenarios~\cite{Xiao:2024,Franke:PRA2021}.}}

{\it Theoretical strategy.-} The atomic system is of the double-$\Lambda$ type, consisting of two ground states $\left|1\right\rangle$ and $\left|2\right\rangle$, and two excited states $\left|3\right\rangle$ and $\left|4\right\rangle$. When states $|2\rangle$ and $\{|3\rangle,|4\rangle\}$ are driven by two strong coupling fields $\Omega_{c,d}$, the weak probe beam $\Omega_p$ driving the transition from $|1\rangle$ to $|3\rangle$ induces a fourth weak signal field $\Omega_s$, forming a FWM process $\left|1\right\rangle\rightarrow\left|3\right\rangle\rightarrow\left|2\right\rangle\rightarrow\left|4\right\rangle\rightarrow\left|1\right\rangle$ \cite{GJY:OE2011}, as shown in Fig. \ref{fig:model}. We note that the detuning $\delta$ of the coupling field $\Omega_d$ can serve as an effective control knob to enhance the CE in FWM by avoiding the multi-photon destructive interference \cite{PRL2003}.
The Hamiltonian is given by
\begin{equation}
   H =  \Omega_{p} \lvert 3 \rangle \langle 1 \rvert + \Omega_{c} \lvert 3 \rangle \langle 2 \rvert + \Omega_{d}\lvert 4 \rangle \langle 2 \rvert + \Omega_{s} \lvert 4 \rangle \langle 1 \rvert + \text{H.c.} +\delta\lvert 4 \rangle \langle 4 \rvert.
\end{equation}
The system dynamics are captured by the density matrix $\rho$ \cite{Lindblad2017}
\begin{equation}
    \frac{\partial\rho}{\partial t}=-i\left[H,\rho\right]+\mathcal{L}_{d}[\rho]+\mathcal{L}_{z}[\rho],
\label{Bloch_exact}    
\end{equation}
where the Lindblad superoperators are
$\mathcal{L}_{d}[\rho]=\sum_{m<n}\left(L_{mn}\rho L_{mn}^{\dagger}-\frac{1}{2}\left\{ L_{mn}L_{mn}^{\dagger},\rho\right\} \right)$ 
with $L_{mn}=\sqrt{\Gamma_{nm}}\left|m\right\rangle \left\langle n\right|$ ($n=3,4;m=1,2$) describing the spontaneous decay from $\left|n\right\rangle$ to $\left|m\right\rangle$ (with rate $\Gamma_{nm}$), and $\mathcal{L}_{z}[\rho]=L_{z}\rho L_{z}^{\dagger}-\frac{1}{2}\left\{ L_{z}L_{z}^{\dagger},\rho\right\}$ with $L_{z}=\sqrt{\gamma}\left(\left|1\right\rangle \left\langle 1\right|-\left|2\right\rangle \left\langle 2\right|\right)$ accounting for the dephasing between two ground states $|1\rangle$ and $|2\rangle$ (with rate $\gamma$).

In order to provide an explicit discussion, we consider the weak-probe approximation where $\Omega_p\ll \Omega_{c,d}$, so that atoms will persistently stay in the initial state $|1\rangle$, satisfying $\rho_{11}\approx 1$, $
\rho_{jj}\approx 0$ ($j\in(2,3,4)$). We retain only the atomic coherence terms associated with $|1\rangle$, yielding a set of reduced equations for density matrix elements
\begin{subequations}
\begin{align}
\frac{\partial}{\partial t}\rho_{41}&=i\left(\Omega_{s}+\Omega_{d}\rho_{21}\right)+\left(i\delta-\frac{\Gamma+\gamma}{2}\right)\rho_{41},  \\
\frac{\partial}{\partial t}\rho_{31}&=i\left(\Omega_{p}+\Omega_{c}\rho_{21}\right)-\frac{\Gamma+\gamma}{2}\rho_{31},   \\
\frac{\partial}{\partial t}\rho_{21}&=i\left(\Omega_{c}^{*}\rho_{31}+\Omega_{d}^{*}\rho_{41}\right)-\gamma\rho_{21},
\end{align}
\label{Bloch}
\end{subequations}
where we have assumed $\Gamma_{nm}=\Gamma/2$ for simplicity. We adopt a zero dephasing rate ($\gamma=0$) to obtain an explicit interpretation with the perfect EIT condition \cite{PRL193604}, while this dephasing is included in the numerical calculations. Using the steady-state assumption, we can directly solve the atomic coherence related to the propagation of probe and signal fields
\begin{subequations}
    \begin{equation}
     \rho_{31}=i\left(-\left|\Omega_{d}\right|^{2}\Omega_{p}-\Omega_{d}^{*}\Omega_{c}\Omega_{s}\right)/D, 
    \end{equation}
    \begin{equation}
    \rho_{41}=i\left(-\Omega_{c}^{*}\Omega_{d}\Omega_{p}+\left|\Omega_{c}\right|^{2}\Omega_{s}\right)/D,
    \end{equation}
    \label{rho31rho41}
\end{subequations}
with all quantities rescaled by the frequency unit $\Gamma$ and 
$D=(1-2i\delta)|\Omega_{c}|^{2}+|\Omega_{d}|^{2}$. In order to study the behavior of the probe ($\Omega_p$) as well as the generated signal ($\Omega_s$) fields propagating in the atomic medium, we use the Maxwell-Bloch equations
\begin{equation}
\frac{\partial\Omega_{p}}{\partial z}=i\frac{\alpha}{2L}\rho_{31},\qquad
\frac{\partial\Omega_{s}}{\partial z}+i\Delta k\Omega_{s}=i\frac{\alpha}{2L}\rho_{41},
\label{Prop}
\end{equation}
in which we include the non-negligible phase mismatch, with its strength measured by $\Delta k$ [see Fig.\ref{fig:model}]. {Note that this phase mismatch arises from the inevitably small angle $\varphi$ used for reducing the leakage from two coupling fields \cite{Finkelstein:2023}}. Here, $L$ denotes the length of atomic medium and $\alpha$ represents OD. We define a dimensionless mismatch parameter $\kappa = \Delta k L$, scaled such that $\kappa \in [0,5]$, corresponding to a relatively small angle $\varphi \lesssim 8^\circ$ for $L = 3.5$ mm.

By inserting Eq.\ref{rho31rho41}(a-b) into the propagation equations (\ref{Prop}), the system forms Schr\"{o}dinger-like coupled equations ($L$ serves as a length unit)
\begin{equation}
    i\frac{\partial}{\partial z}\left(\begin{array}{c}
\Omega_{p}\\
\Omega_{s}
\end{array}\right)=  \mathcal{M}\left(\begin{array}{c}
\Omega_{p}\\
\Omega_{s}
\end{array}\right)
\label{propa_theta}
\end{equation}
where 
\begin{equation}
    \mathcal{M}=\left(\begin{array}{cc}
\zeta\sin^{2} \theta & -\zeta\sin (2\theta)/2\\
-\zeta\sin (2\theta)/2 & \zeta\cos^{2} \theta+\kappa
\end{array}\right)
\label{H}
\end{equation}
with the coefficient given by
\begin{equation}
\zeta=\frac{\alpha}{2} \left( \frac{2 \delta \cos^2 \theta -i}{1 + 4 \delta^2 \cos^4 \theta} \right). 
\label{zeta}   
\end{equation}  
Remarkably, we define a new ratio $\Omega_c/\Omega_d=\cot \theta$, indicating that propagation depends only on the ratio of the two coupling fields rather than their absolute values. In principle, $\Omega_{c}$ and $\Omega_{d}$ 
can vary arbitrarily in space, provided that the dark eigenmode $\psi_D$ is maintained (see Eq.\ref{dark}). Additionally, the spontaneous loss, incorporated through the non-Hermitian matrix $\mathcal{M}$
via the imaginary part of $\zeta$, scales as $\propto e^{-\alpha/2}$ for both $\Omega_p$ and $\Omega_s$, decreasing exponentially with increasing $\alpha$. Hence, a dense medium with a large $\alpha$ can achieve near-unity signal conversion \cite{Shui:PRA2023}. In what follows we consider three protocols for controlling efficient FWM. Protocol I (‘{\it Constant-Optimization}’) uses fixed, position-independent parameters; Protocol II (‘{\it Unoptimized Dark-State}’) employs simple linearly-varying coupling fields based on dark state without further optimization; and Protocol III (‘{\it Spatially-Dependent Detuning}’) introduces a spatially-varying detuning, which significantly improves CE under phase mismatch.

\begin{figure}
    \centering
    \includegraphics[width=0.3\textwidth]{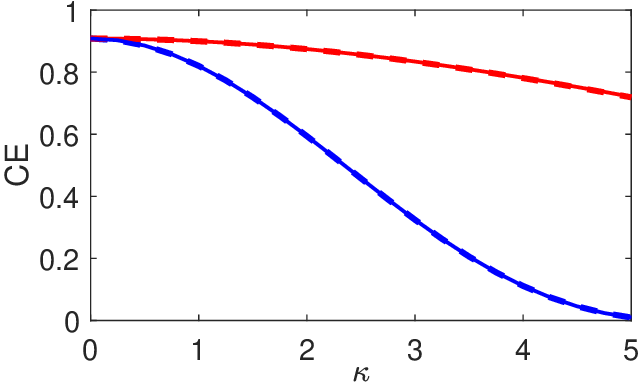}
    \caption{\justifying Comparison of CE as a function of phase mismatch $\kappa$, with optimal parameters ($\theta_{opt}(\kappa),\delta_{opt}(\kappa)$) (red, Protocol I) and fixed parameters ($\theta_{opt}(0),\delta_{opt}(0)$) (blue). The realistic behavior of CE solved from Eq.(\ref{Bloch_exact}) is shown by the solid lines. While the theoretical predictions shown by dashed lines are plotted using the computed CE function in Eq.(\ref{CE}). Other parameters are  $\alpha=200$, $\gamma=0$, $\Omega_p=0.03$, $\Omega_d=1.5$, $\Omega_c=\Omega_d\cot\theta$ is $\kappa$-dependent.}
    \label{Fig:surfplot}
\end{figure}

{\it {Protocol I: Constant-Optimization.}} In general, Eq.(\ref{propa_theta}) can be solved exactly under constant parameters. We define the conversion efficiency as $CE=|\Omega_s(z=1)|^2/|\Omega_{p0}|^2$ which quantifies the performance of signal transmission. An explicit form is given by
\begin{equation}
    CE=\frac{|\zeta\sin(2\theta)|^2}{4\lvert{\lambda_{+}-\lambda_{-}}\rvert^2} \lvert e^{\lambda_{+}}-e^{\lambda_{-}}\rvert^2 
\label{CE}
\end{equation}
where $\lambda_{\pm}=\frac{1}{2}\left(\zeta+\kappa\pm\sqrt{\zeta^{2}+\kappa^{2}+2\kappa\zeta\cos (2\theta)}\right)$
are the eigenvalues of $\mathcal{M}$. 
To maximize CE, in Protocol I we introduce the {\it constant-optimization} method, where the diagonal energy shift of matrix $\mathcal{M}$ is set to vanish and a $\pi$-pulse coupling is enforced via $\int_0^1 |\zeta\sin (2\theta)|dz=\pi$, based on which we obtain an optimal set of $\kappa$-dependent constants ($\theta,\delta$) for a best CE: 
\begin{equation}
 \theta_{opt}=\frac{\pi}{4}+\frac{1}{2}\text{arctan}(\kappa/\pi),
 \delta_{opt}=\frac{\alpha}{2\pi^2}(\kappa+\sqrt{\kappa^2+\pi^2})
 \label{para_opt}
\end{equation}
For the special case of perfect phase matching $\kappa=0$, one gets $(\theta_{opt},\delta_{opt})=(\frac{\pi}{4}, \frac{\alpha}{2\pi})$ yielding the maximal CE  
\begin{equation}
   CE= \frac{1}{4}(1+e^{-\frac{2\pi^2}{\alpha}})^2.
\end{equation} 
At a large OD, e.g., $\alpha=200$, this gives CE=0.911, comparable to other FWM protocol \cite{Chen:OL2021}. For a moderate OD ($\alpha=50$), CE decreases to 0.700 with phase-matching.
However, any nonzero phase-mismatch $\kappa\neq 0$ would be detrimental to the high-efficiency generation of FWM signals.

Figure \ref{Fig:surfplot} shows the CE impacted by $\kappa$ in a double-$\Lambda$ FWM system. If the parameters are fixed at their optimal phase-matched values $(\theta,\delta)=(\frac{\pi}{4},\frac{\alpha}{2\pi})$ (blue curves), CE drops rapidly with increasing $\kappa$, reaching $CE\approx 0. 01$ at $\kappa =5$. In contrast, the analytic solution of Eq.(\ref{para_opt}) developed in Protocol I (red curves) yields substantial improvement: for any given $\kappa$, setting the coupling fields in the ratio $\Omega_c = \Omega_d\cot\theta_{opt}$ together with the $\kappa$-dependent detuning $\delta_{opt}$ significantly enhances CE, reaching 0.72 for $\kappa=5$. 
The full numerical solutions (solid) based on the density matrix equation (\ref{Bloch_exact}) perfectly validate the scheme accuracy (dashed). Thus, this {\it Constant-Optimization} strategy can partly compensate for phase mismatch, but requires a precise knowledge of $\kappa$, which is unrealistic in experiments, motivating a more robust space-dependent optimization.

\begin{figure}
    \centering
    \includegraphics[width=0.4\textwidth]{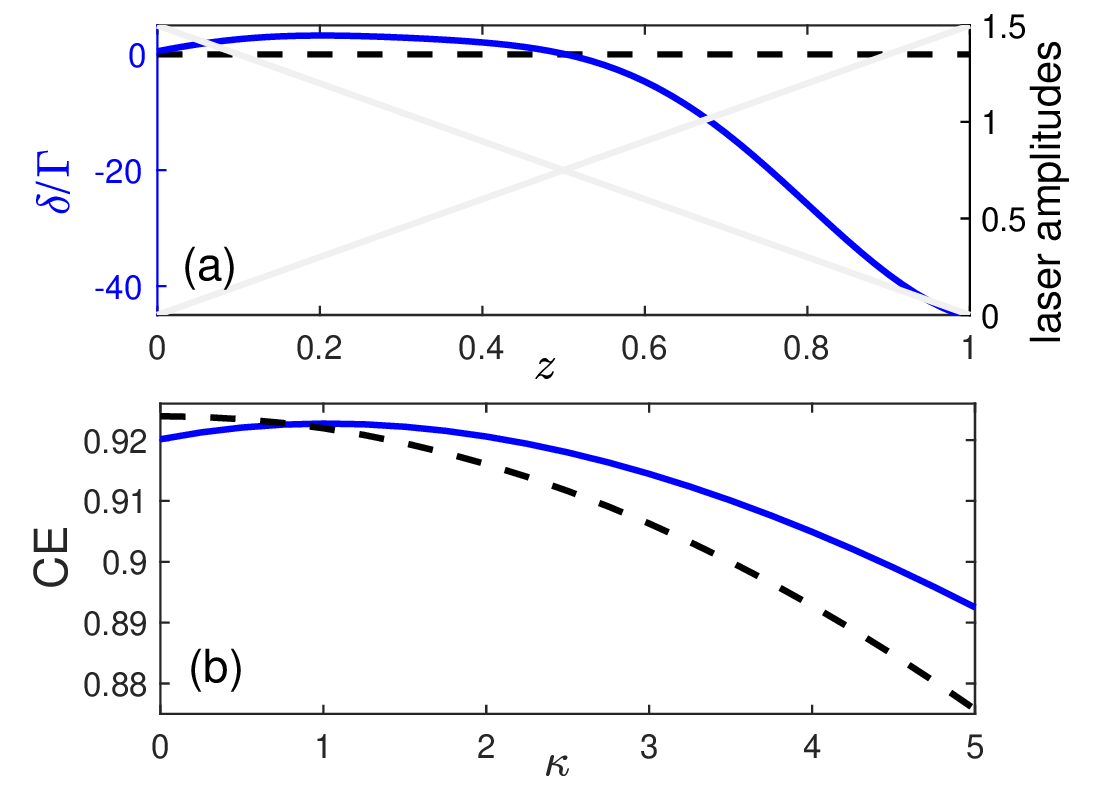}
    \caption{\justifying Performance of the FWM scheme  as a function of phase mismatch $\kappa$ under different optimized waveforms $\delta(z)$. The curves comparably represent the cases of $\delta(z)=0$ (Protocol II, dashed) and $\delta=\delta(z)$ (Protocol III, solid) where all coefficients are $(a_0,a_1,a_2,a_3,b_1,b_2,b_3)=(-13.55,20.15,6.17   -8.93,-0.91,2.86,1.45)$. (a) and (b) correspond to the detuning waveforms and the calculated CE values, respectively. The light gray curves in (a) indicate the unchanged adiabatic amplitudes of $\Omega_{c}$ and $\Omega_d$. Relevant parameters are $\alpha=200$, $\Omega_{p}=0.03$, $\gamma=10^{-4}$, $\Omega_{c0}=\Omega_{d0}=1.5$.}  
    \label{Fig:Fig3}
\end{figure}


{\it {Protocols II and III: Space-dependent optimization.}} 
To further improve the CE, especially in the case of a large phase mismatch, we introduce  two orthogonal base vectors
\begin{equation}
\psi_{D}=\left(\cos\theta,  \sin\theta\right)^{T},
\qquad
\psi_{B}=\left(\sin\theta,   -\cos\theta\right)^{T},
\label{dark}
\end{equation}
which coincide with the eigenstates of $\mathcal{M}$ for $\kappa=0$, corresponding to eigenvalues $\lambda_+=0$ and $\lambda_-=\zeta$. In the phase-matched case, $\psi_D$ forms a dark state, fully protected from spontaneous dissipation arising from the imaginary part of $\zeta$. Ideally, with $\theta(z=0)=0$, the system can adiabatically evolve along the dark state $\psi_D$ to achieve a complete transfer from $\Omega_p$ to $\Omega_s$ at $z=1$, requiring $\theta(z=1)=\pi/2$. To test the above adiabatic control of $\theta(z)$, we introduce spatially varying amplitudes for the two coupling fields, analogous to the counter-intuitive pulse sequence used in adiabatic passage \cite{Moses:PRL2020}
. Here $\Omega_{c,d}(z)$ are simply linearly modulated at lower technical cost, following the implementation in \cite{Wu:OE2024} 
\begin{equation}
   \Omega_c(z)= \Omega_{c0} (1- z),
   \qquad
   \Omega_d(z)= \Omega_{d0} z
   \label{line}
\end{equation}
 where $\Omega_{c0},\Omega_{d0}$ denote the maximal amplitudes. This configuration allows global optimization to be performed solely through the detuning $\delta(z)$, while maintaining the adiabatic evolution along the dark state. Notably, the detuning waveform required to achieve high CE is not unique, providing a degree of freedom that can be exploited either to maximize the average CE over a broad range of $\kappa$ or to optimize CE for a specific large $\kappa$.

We utilize a numerical GA to determine an optimal waveform $\delta(z)$ that can achieve a higher CE even for relatively large $\kappa$. We take the average value $F=\sum _{i=1}^N CE(\kappa_i)/N$ as the target and minimize the cost function $1-F$ under sufficient samplings $N=20$ for $\kappa_i= i/4, i=0,...,N$. We first assume the Fourier series form  
\begin{equation}
   \delta(z)=a_0+\sum_{n=1}^{3}[a_n\cos(n\pi z)+b_n\sin(n\pi z)] 
   \label{four}
\end{equation}
The detuning waveform is a smooth function in space, facilitating experimental implementation. Note that the coupling amplitudes $\Omega_{c,d}(z)$ are kept unchanged as linear modulation.

To demonstrate the improved robustness of FWM protocol against the phase-mismatch effect, we first calculate the CE without optimization by choosing $\delta=0$ labeled as {Protocol II: {\it Unoptimized Dark-State}.} 
Excitingly, thanks to the presence of a dark mode $\varphi_D$ that adiabatically protects the system from spontaneous loss, 
the CE values displayed in Fig.\ref{Fig:Fig3}b by the dashed line can achieve a dramatically increased insensitivity to $\kappa$. Even at $\kappa=5$, the CE remains as high as 0.876, outperforming the maximum 0.72 achieved in Protocol I (see Fig.\ref{Fig:surfplot}). {As turning to the Protocol III: {\it Spatially-Dependent Detuning}}, labeled by the solid line, we can obtain stronger robustness against the change of $\kappa$. In Fig.\ref{Fig:Fig3}b, the CE increases to 0.893 for $\kappa=5$, though at the cost of slightly reduced, yet still high CE values for small mismatch $\kappa  \leq 1$. Because our numerical algorithm requires the maximization of average CE value within $\kappa\in[0,5]$, it tends to improve the overall insensitivity of scheme performance to the phase-mismatch effect. Specifically, in Protocol III the optimized $\delta(z)$ function needs to be far from resonance after the adiabatic transfer at $\Omega_c=\Omega_d$, aiming to compensate for the large phase mismatch $\kappa$ by employing a largely negative detuning $\delta$ ($\approx -45.5\Gamma$) see Fig. \ref{Fig:Fig3}a.

 {\it Mitigating larger phase mismatch.-} We demonstrate that the spontaneous loss due to a finite OD can be decreased by applying an off-resonant detuning, as in Protocol III. A large negative detuning can also mitigate the significant (positive) phase-mismatch effect in practice. A natural question at this stage is whether such space-dependent optimal control of $\delta(z)$ can further improve FWM performance under larger phase mismatch and moderate OD values. To address this, we examine CE at $\kappa = 15$ (corresponding to the separated angle $\varphi \approx 13.6^{\circ}$) with $\alpha=50$. {{This extreme mismatch case, though beyond typical experimental conditions, is intentionally considered to highlight the robustness and potential of the proposed protocol under demanding scenarios.}} This choice only gives CE $\approx 0.019$ based on Protocol I with optimal constants $(\theta_{opt},\delta_{opt})=(0.467\pi,76.82)$. Switching to Protocol II improves CE to 0.132, but the simple spatial-light modulation remains insufficient to overcome the severe energy loss caused by large phase mismatch.

 \begin{figure}
    \centering
    \includegraphics[width=0.4\textwidth]{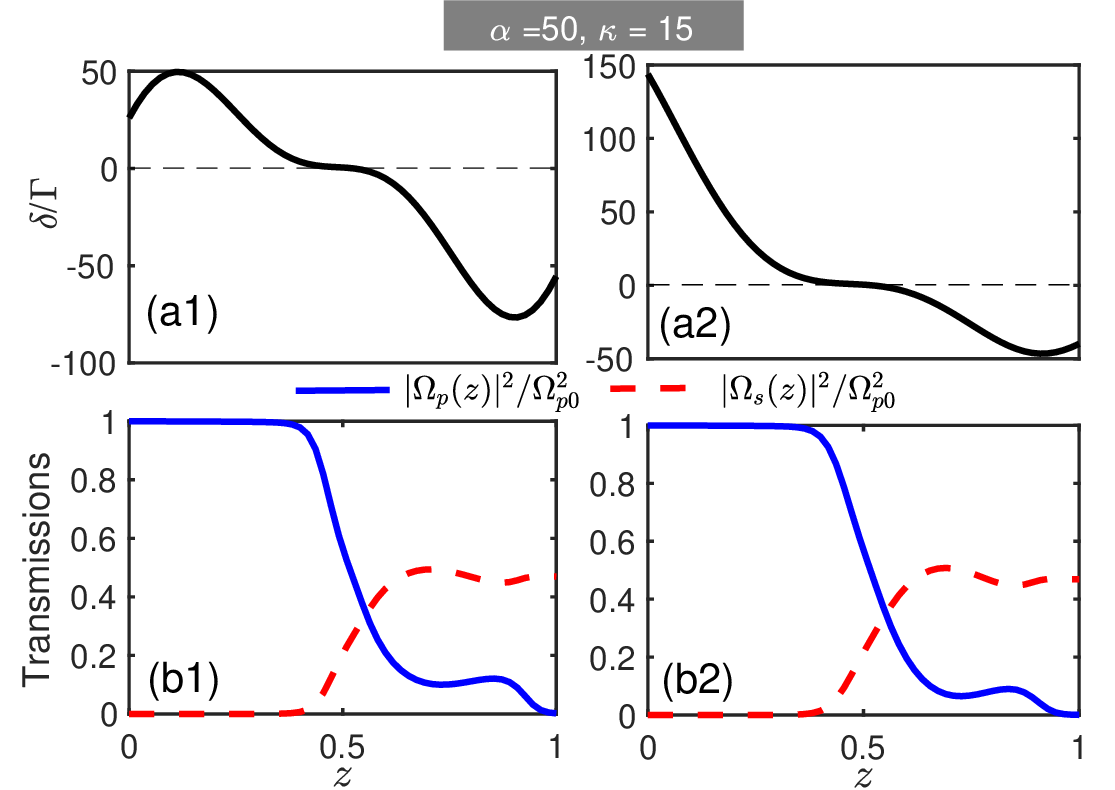}
    \caption{\justifying (a1-2) The optimized waveforms of coupling detuning function. (a1) $\delta(z)$ takes the Fourier-series form with coefficients $(a_0,a_1,a_2,a_3,b_1,b_2,b_3)$=(-1.73,-2.32,-9.08,-13.04,67.86,43.09,1.74). (a2) A different waveform uses Bernstein polynomial where $\delta(z)=\sum_{n=0}^{7}a_{n}C_{7}^{n}z^{n}(1-z)^{7-n}$ with coefficients $(a_0,a_1,a_2,...,a_7)$=(143.76,71.73,-42.35,-53.98,111.60,-59.90,-59.94,-39.85). Here $C_{7}^{n}$ denotes the binomial coefficient. (b1-2) The corresponding transmissions of the probe (solid) and signal (dashed) fields. Here $\alpha=50,\kappa=15$ and other parameters are the same as in Fig.\ref{Fig:Fig3}.}  
    \label{Fig:Fig4}
\end{figure}

 In order to achieve a highly efficient conversion for larger mismatch strength in a more practical setting, Protocol III employs GA to search for an optimized detuning waveform $\delta(z)$ at a specific mismatch value $\kappa=15$, using $1-CE(\kappa=15)$ as the cost function. This ensures maximization of CE around the target large phase mismatch. Note that this may not be the best optimized waveform when $\kappa<15$.
The resulting optimized waveforms of $\delta(z)$ are plotted in Fig. \ref{Fig:Fig4}.
For comparison, we also employ the Bernstein polynomial as an alternative choice. As observed in Fig. \ref{Fig:Fig4}(a1-a2), the optimized $\delta(z)$ functions in both cases exhibit large off-resonance at the boundaries ($z=0$ or $1$) to suppress spontaneous dissipation loss. Meanwhile, to obey the adiabatic transfer along the dark eigenmode $\varphi_D$ ensured by $\Omega_c$ and $\Omega_d$, 
both $\delta(z)$ functions approach near-resonance around $z=0.5$, where the probe-to-signal field conversion occurs. The corresponding space-dependent transmission of the probe and signal fields is depicted in Fig. \ref{Fig:Fig4}(b1–b2). Remarkably, both implementations under Protocol III achieve an enhanced CE of approximately $ 0.469$, significantly outperforming Protocol I (0.019) and Protocol II (0.132), highlighting the effectiveness of space-dependent detuning optimization. This also demonstrates that the optimization procedure is robust with respect to the choice of functional form, motivating further exploration of different detuning waveforms for experimental implementation.

{\it Conclusion.-} We present a novel scheme to mitigate the detrimental effects of phase-mismatch in a non-resonant FWM system using optimal control strategy. It is demonstrated that when the two coupling fields are maintained as fixed linear spatial modulations, an optimized waveform for a single coupling detuning can strongly suppress the spontaneous decay loss from two excited states, leading to a substantial enhancement in the CE even for the large phase-mismatch regime. Our Protocol III with its carefully designed detuning provides superior robustness against variations in mismatch while preserving high-efficiency signal transmission, a key requirement for all-optical quantum information processing such as in frequency conversion involving structured light modes where high efficiency is crucial \cite{Offer2018}.
These features are expected to significantly relax the technical constraints on phase-matching conditions, motivating further exploration in diverse scenarios \cite{acoustic,biphoton}.

\begin{backmatter}
\bmsection{Funding} National Natural Science Foundation of China (12174106, 11474094, 12104308); Natural Science Foundation of Chongqing Municipality (CSTB2024NSCQ-MSX1117); Research Council of Lithuania (LMTLT) (S-ITP-24-6).

\bmsection{Disclosures} The authors declare no conflicts of interest.

\bmsection{Data Availability Statement} Data underlying the results presented in this paper are not publicly available at this time but may be obtained from the authors upon reasonable request.

\end{backmatter}

\bibliography{sample}







\end{document}